\documentclass[11pt, oneside]{article}   	
\usepackage[margin=1in]{geometry}                		
\usepackage{graphicx,bbm,subfigure}				
\usepackage{amssymb}
\usepackage{amsmath,cite}
\usepackage{xcolor}
\usepackage{bm}
\usepackage[colorlinks,citecolor=blue]{hyperref}
\usepackage{cancel}

\newcommand{\ket}[1]{\left| #1\right\rangle}

\def\x{\mathbf{x}}
\def\y{\mathbf{y}}
\def\q{\mathbf{q}}
\def\p{\mathbf{p}}

\def\s{\mathbf{s}}
\def\P{\mathcal{P}_{\text{staggered}}}
\def\lcm{\text{lcm}}
\def\ta{\tilde{a}}
\def\tn{\tilde{n}}
\def\tvarphi{\tilde{\varphi}}
\def\W{\widehat{W}}

\usepackage[margin=1.5cm,font=small,labelfont=bf]{caption}

\def\RR{\mathbb R}
\newcommand{\ZZ}{\mathbb{Z}} 

\title{\bf Canonical quantization of lattice Chern-Simons theory}
\author{Theodore Jacobson$^1$\footnote{tjacobson@physics.ucla.edu}\,\, and Tin Sulejmanpasic$^2$\footnote{tin.sulejmanpasic@durham.ac.uk} \\ \\
{\it\small $^1$Mani L. Bhaumik Institute for Theoretical Physics, Department of Physics and Astronomy,}\\
{\it\small University of California, Los Angeles, CA 90095, USA} \\
{\it\small $^2$Department of Mathematical Sciences, Durham University, DH1 3LE Durham, UK}
}
\date{\today}							

\begin{document}
\maketitle
\thispagestyle{empty}

\abstract{
We discuss the canonical quantization of $U(1)_k$ Chern-Simons theory on a spatial lattice. In addition to the usual local Gauss law constraints, the physical Hilbert space is defined by 1-form gauge constraints implementing the compactness of the $U(1)$ gauge group, and (depending on the details of the spatial lattice) non-local constraints which project out unframed Wilson loops. Though the ingredients of the lattice model are bosonic, the physical Hilbert space is finite-dimensional, with exactly $k$ ground states on a spatial torus. We quantize both the bosonic (even level) and fermionic (odd level) theories, describing in detail how the latter depends on a choice of spin structure.

}

\newpage 
{\hypersetup{linkcolor=black}
\tableofcontents
}

\section{Introduction}

Chern-Simons (CS) theory is the quintessential topological quantum field theory (TQFT), with numerous applications and appearances throughout condensed matter and high-energy physics. Even the simplest abelian CS theory plays a prevalent role in the study of anomalies, level/rank and boson/fermion dualities, defects in higher-dimensional theories, and in the quantum field theories of spin liquids and the fractional quantum hall effect. 

The literature is host to a multitude~\cite{Frohlich:1988qh,Kavalov:1989kg,Eliezer:1991qh,Eliezer:1992sq,Luscher:1989kk,Muller:1990xd,Sun:2015hla,DeMarco:2019pqv} of distinct formulations of Chern-Simons theory on the lattice, whose variety reflects the lack of universality familiar from continuum quantum field theory. In a previous paper~\cite{Jacobson:2023cmr} we realized $U(1)_k$ CS theory as a lattice gauge theory on a Euclidean spacetime lattice using the modified Villain formalism~\cite{Villain:1974ir,Sulejmanpasic:2019ytl}. This framework can be used to endow lattice models with key features at finite lattice spacing which in a conventional discretization would only emerge in the continuum limit.\footnote{See e.g. Refs.~\cite{Gorantla:2021svj,Choi:2021kmx,Anosova:2022cjm,Anosova:2022yqx,Yoneda:2022qpj,Fazza:2022fss,Cheng:2022sgb,Thorngren:2023ple,Seifnashri:2023dpa,Berkowitz:2023pnz,Chen:2019mjw,Han:2021wsx,Han:2022cnr} for applications of this formalism, and especially Refs.~\cite{Chen:2019mjw,Han:2021wsx,Han:2022cnr} which employ the modified Villain approach to `doubled' CS theory on the lattice. } The present paper follows up on our previous work~\cite{Jacobson:2023cmr} and is dedicated to the canonical quantization of abelian CS theory on a spatial lattice within the Villain Hamiltonian~\cite{Fazza:2022fss} approach. Given the long list of works on this subject, one may wonder what new insights can be gained by revisiting this problem. We find that the lattice Hamiltonian formulation provides useful perspectives on certain crucial features of CS theory which were preserved by our spacetime discretization, such as compactness of the gauge group, level quantization, the framing of Wilson lines, the anomalous 1-form symmetry, and more. 
 
One of the main advantages of working in the Hamiltonian formulation is that we have direct access to the Hilbert space. The building blocks of our lattice CS theory are gauge fields on links and discrete integer fields on plaquettes, which are each associated with infinite-dimensional Hilbert spaces. However, it is well-known that the Hilbert space of CS theory in the continuum has a finite dimension equal to $k^g$, where $k$ is the level and $g$ is the genus of the spatial manifold. This apparent discrepancy is resolved in the following way: the physical Hilbert space of our lattice model is actually a projection of the infinite-dimensional Hilbert space down to a finite-dimensional constrained Hilbert space. There are three types of constraints: the local Gauss law implementing `small' gauge transformations, a discrete 1-form Gauss law implementing `large' gauge transformations, and a non-local constraint which projects out unframed Wilson lines. Together, these constraints imply that the only non-trivial operators are framed, topological Wilson loops, whose algebra results in a physical Hilbert space whose dimension matches that of the continuum theory exactly. 

In addition to repeating much of the analysis of Ref.~\cite{Jacobson:2023cmr} in the canonical formalism, we go further and discuss how to formulate the odd-level CS theories, which in the continuum are spin-TQFTs requiring a choice of (and exhibiting a dependence on) spin structure. Using the method of `fermion condensation' well-known in the TQFT literature~\cite{Gaiotto:2015zta,Bhardwaj:2016clt}, we construct a consistent Hilbert space for the odd-$k$ CS theories and show that Wilson loops have the spin structure-dependence expected from the continuum. In our view this comprises a significant step towards establishing nonperturbative lattice-level fermion-boson dualities, and of placing the decades-old idea of flux-attachment~\cite{PhysRevLett.48.1144,Polyakov:1988md} on equal footing with particle-vortex duality, which originated on the lattice~\cite{Peskin:1977kp,Dasgupta:1981zz}. For similar work in this direction but without explicit Chern-Simons terms on the lattice, see Refs.~\cite{Chen:2017lkr,Chen:2018vmz}.

\begin{figure}[h] 
   \centering 
\includegraphics[width=.7\textwidth]{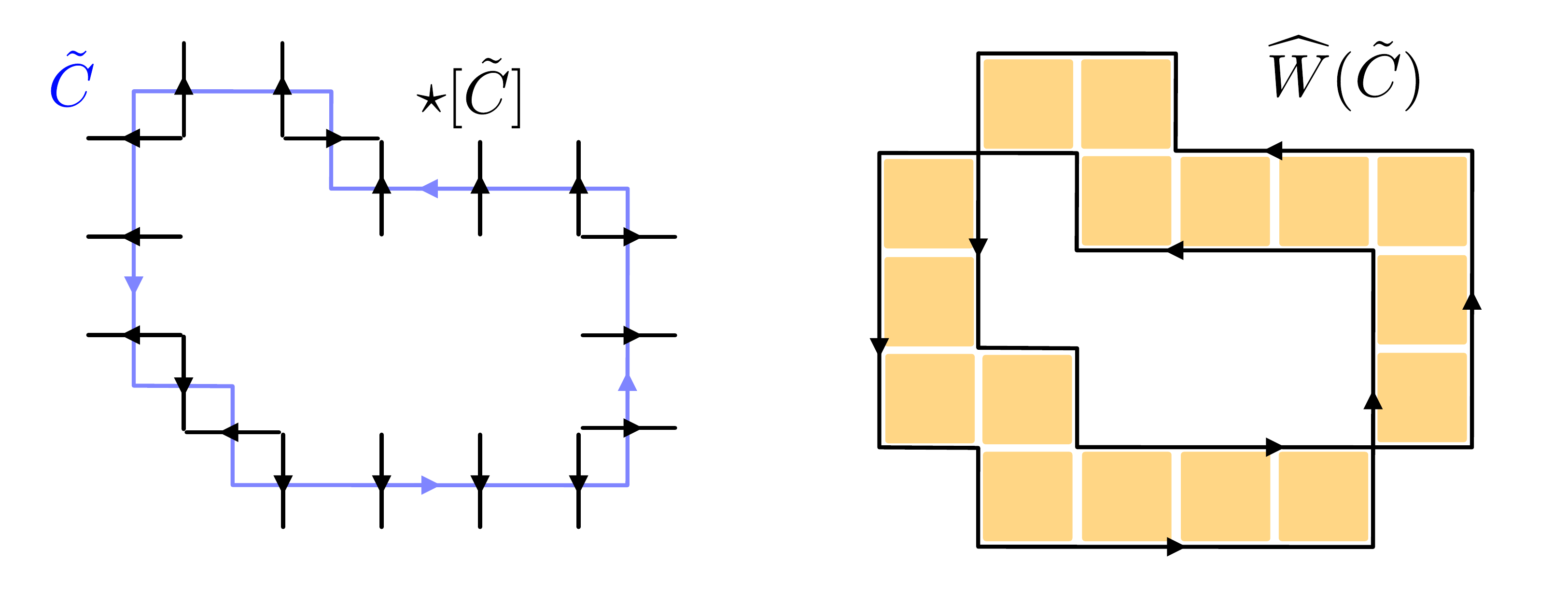} 
\caption{The physical observables in the compact $U(1)$ Chern-Simons theory on the lattice are framed Wilson lines, or ribbon operators. The ordinary Wilson lines at the two edges of the ribbon are connected by a surface operator built from integer-spectrum Villain fields, such that the ribbons as a whole are gauge-invariant and topological --- they only depend on the homology class of the line $\tilde C$ on the dual lattice which defines their support. }
\label{fig:compact_framed}
\end{figure}

Much of the literature on lattice CS theory was concerned by the presence of additional gauge field zero modes in the most obvious discretization of the continuum CS term. As summarized in Ref.~\cite{Jacobson:2023cmr}, these zero modes turn out to be crucial, rather than detrimental, as they capture the need for framing --- i.e. the fact that the physical Wilson lines in CS theory are ribbons rather than lines~\cite{Witten:1988hf}. As mentioned above and discussed in more detail below, these zero modes are associated with additional constraints which must be imposed on the Hilbert space which project out all unframed Wilson lines. The need for additional constraints was noticed in Refs.~\cite{Eliezer:1991qh,Eliezer:1992sq} but deemed unphysical --- these authors gave an alternative discretization of (continuum time) abelian CS theory without zero modes (see Ref.~\cite{Sun:2015hla} for the generalization to an arbitrary graph). Roughly speaking these constructions correspond to choosing a framing in the time direction. In contrast, we take the view that the zero modes are a crucial part of the quantization procedure, and our resulting lattice theory makes manifest certain global aspects which are obscured in these alternative formulations. 

We begin in Sec.~\ref{sec:non_compact} by discussing the non-compact CS theory with gauge group $\RR$. The zero modes and non-local constraints leading to framing are already present in this simple starting point. In Sec.~\ref{sec:compact} we introduce discrete Villain fields to gauge a subgroup of the 1-form symmetry of the non-compact model, making the gauge group $U(1) = \RR/2\pi \ZZ$. Without additional ingredients, consistency of the gauging procedure quantizes the level $k$ to be an even integer. Canonically conjugate to the Villain field is a compact scalar which plays the role of the monopole operator, and carries electric charge $k$. We discuss the algebra of Wilson loops, its relation to the 't Hooft anomaly of the $\ZZ_k$ 1-form symmetry, and the implications for ground state degeneracy. Then in Sec.~\ref{sec:fermionic_CS} we introduce a pair of Majorana fermions on each plaquette, and use them to consistently quantize the odd level theories in a way which depends on the spin structure of the spatial lattice. We conclude in Sec.~\ref{sec:conclusions} with directions for future work.

\section{Warmup: the non-compact theory}
\label{sec:non_compact} 

Consider Hamiltonian lattice gauge theory~\cite{Kogut:1974ag} formulated on a rectangular spatial lattice $\Lambda$ where time is real, continuous, and non-compact. We work on a spatial lattice with $L_1$ and $L_2$ sites in each direction and periodic boundary conditions, so that the spatial topology is that of a torus. We denote the sites, links, and plaquettes by $x, \ell, p$, sometimes using vectorial notation so that $x = \x$ is the site at $\x$, $\ell = (\x,i)$ is the link starting at $\x$ pointing in the $i$ direction, and $p = (\x,12)$ is the plaquette whose lower left corner is at $\x$. Throughout the paper we use the language of cochains and lattice differential forms. Functions on sites, links, and plaquettes are called 0-, 1-, and 2-cochains, with $C^i(R)$ denoting the set of $R$-valued $i$-cochains. We make heavy use of cup products, which are the lattice analogs of wedge products in de Rham cohomology. We stick to the conventions of Refs.~\cite{Sulejmanpasic:2019ytl,Jacobson:2023cmr} --- see those references for further details.\footnote{We will repeatedly use the following crucial formulas --- one is the analog of the Leibniz rule, or `summation by parts'
\begin{equation}
d(\alpha\cup\beta) = d\alpha\cup\beta + (-1)^p\, \alpha\cup d\beta,
\end{equation}
and the other is an identity capturing the lack of super-commutativity of cup products:
\begin{equation} \label{eq:cup_commutativity} 
\alpha \cup \beta - (-1)^{pq} \, \beta\cup \alpha = (-1)^{p+q+1}\left[ d(\alpha\cup_{1}\beta)- d\alpha\cup_{1}\beta - (-1)^p\, \alpha \cup_{1}\beta\right]\,,
\end{equation}
where $\alpha, \beta$ are $p-$ and $q-$cochains respectively. Here $\cup_1$ is the cup-1 product --- the $\cup_i$ products vanish unless $i \le p,q$. For general formulae for and applications of (higher) cup products on the (hyper) cubic lattice, see Ref.~\cite{Chen:2021ppt}.\label{fn:cup_identity} } 

Our starting point is a real link field (i.e. a real-valued 1-cochain) $a_\ell$ together with a real field $(a_0)_x$ on sites (i.e. a real-valued 0-cochain) representing the spatial and time components of the non-compact $\RR$ gauge field. We begin with the lattice action
\begin{equation}\label{eq:action}
S=\frac{k}{4\pi} \int dt \sum_{p\in \Lambda} \Big[ (a\cup (\dot a-da_0))_p-(a_0\cup da)_p\Big]\;,
\end{equation}
with $k$ an arbitrary real number. Summing by parts, the above can also be written as
\begin{equation}
S=\frac{k}{4\pi} \int dt \sum_{p\in \Lambda} \Big[ (a\cup \dot a)_p-(da\cup a_0+a_0\cup da)_p\Big]\;.
\end{equation}
In the standard lattice notation it is
\begin{equation}
S=\frac{k}{4\pi} \int dt \sum_{\x}\Big[\epsilon^{ij} a_{\x,i}\, \dot a_{\x+\hat i,j}-(a_0)_{\x} \left( (da)_{\x-\s,12}+ (da)_{\x,12} \right)\Big]\;,
\end{equation}
where we have defined $\s \equiv \hat 1 + \hat 2$ and the sum over $i,j=1,2$ is implicit. 
Now consider the gauge transformation
\begin{equation}
a\rightarrow a+d\lambda\,, \ a_0\rightarrow a_0+\dot \lambda\,.
\end{equation}
where $\lambda$ is a real valued 0-cochain with arbitrary time dependence. The action is invariant under the above gauge transformation (which is easiest to see in the cup-product formulation). Explicitly, 
\begin{multline} \label{eq:small_gauge_variation} 
\Delta S = \frac{k}{4\pi}\int dt \sum_\Lambda \left[ d\lambda \cup \dot a + a \cup d\dot\lambda + d\lambda \cup d\dot\lambda - da\cup\dot\lambda -\dot\lambda \cup da \right] \\
= \frac{k}{4\pi}\int dt \sum_\Lambda \left[\partial_t(d\lambda \cup a)- d(\dot\lambda \cup a) - d(a\cup\dot\lambda) + d(\lambda \cup d\dot\lambda) \right]\,,
\end{multline}
using the Leibniz rule. Since we take time to be non-compact and the spatial lattice has no boundary, the gauge variation vanishes.  

From the above action we see that the canonical momentum of $a_0$ is zero, forcing a Gauss constraint\footnote{The Hamiltonian would be proportional to $a_0(\dots)$, and since the canonical momentum of $a_0$ has to commute with the Hamiltonian, we must have that $(\dots)=0$.}
\begin{equation}\label{eq:Gauss}
\mathcal G_{\x} = (da)_{\x-\s,12} + (da)_{\x,12}=0 \,.
\end{equation}
Note that this is not what we expect in the CS theory in the continuum, where the flux of $a$ is set to zero, which renders Wilson lines topological. The reason for the above constraint instead of $da=0$ is the lack of graded commutativity of the cup product.

We proceed with the quantization. We write the action in terms of the momentum space fields
\begin{equation}
a_{\x,i}=\frac{1}{\sqrt{V}}\sum_{ \p } \ta_{\p,i}\, e^{i\p\cdot \x}\;,
\end{equation} 
where $\ta_{\p,i}=\ta_{-\p,i}^\dagger$, and $V = L_1L_2$ is the spatial volume. The action becomes 
\begin{equation}
S = \frac{k}{4\pi}\int dt \sum_{\p} \epsilon_{ij}\, \ta_{\p,i}\dot \ta_{-\p,j}\, e^{-i\p_i} + \cdots ,
\end{equation}
where the dots indicate terms that vanish when the Gauss constraint is imposed. We can integrate by parts to write this as
\begin{equation} \label{eq:momentum_space_action} 
S = \frac{k}{8\pi}\int dt \sum_{\p} \epsilon_{ij} ( e^{-ip_i} + e^{ip_j})\, \ta_{\p,i}\dot \ta_{-\p,j}  = \frac{1}{2}\int dt \sum_{\p}  K(\p)_{ij}\, \ta_{\p,i} \dot \ta_{-\p,j} \,,
\end{equation}
where
\begin{equation}
K(\p)_{ij}=\frac{k}{4\pi}\begin{pmatrix}
0 & e^{-ip_1}+e^{i p_2}\\
-e^{ip_1}-e^{-ip_2} & 0
\end{pmatrix} = \frac{k}{4\pi}\epsilon_{ij}(e^{-i p_i} + e^{i p_j}) 
\end{equation}
obeys the antisymmetry property $K(\p)_{ij}=-K(-\p)_{ji}$. Since the action is first order in time derivatives, we can simply read off that the conjugate momentum to $\ta_{\p,i}$ is $K(-\p)_{ji} \ta_{-\p,j} = - K(\p)_{ij} \ta_{-\p,j}$.\footnote{This conclusion also follows from a more careful analysis using Dirac brackets. In this approach one has second-class constraints relating momenta and coordinates $\Pi_{\p,i} = -\tfrac{1}{2}K(\p)_{ij}\ta_{\p,j}$, along with the Gauss law constraint. See Appendix~\ref{sec:first_order} for more details, along with an example of this approach in a simpler setting.} The commutation relations would be obtained by inverting $K(\p)_{ij}$. However, $K(\p)_{ij}$ vanishes on a special set of momenta, namely $\p$ such that $p_1 + p_2  = \pi$ mod $2\pi$, making it singular. As we will repeatedly refer to these special momentum modes throughout the paper, it will be convenient to define
\begin{equation}
\P = \{\mathbf{p} \, | \, p_1 + p_2 = \pi \text{ mod } 2\pi \}\,.
\end{equation}
The size of this set depends sensitively on the lengths of the torus, and is equal to
\begin{equation}
|\P| = \begin{cases}
\gcd(L_1,L_2) & \, \text{ if $\lcm(L_1,L_2)$ is even,} \\
0 & \, \text{ if $\lcm(L_1,L_2)$ is odd.}
\end{cases}
\end{equation}
If $\p\in\P$, the vanishing of $K(\p)_{ij}$ simply means that the conjugate momentum for $\ta_{\p,i}$ vanishes, 
\begin{equation} \label{eq:staggered_constraint} 
\Pi_{\p,i} = 0, \ \p \in \P\,.
\end{equation}
In the following we will refer to this constraint as the `framing constraint,' for reasons which will become clear soon.  

We can proceed with the modes for which $\p \not\in \P$. For these modes, $K(\p)_{ij}$ is invertible, 
\begin{equation}
(K(\p)^{-1})_{ij} = -\frac{1}{2}\left(\frac{4\pi}{k}\right)^2\frac{1}{1+\cos(p_1+p_2)}K(\p)_{ij}\,,
\end{equation}
so the commutation relations are given by
\begin{equation} \label{eq:bcommutator} 
[\ta_{\p,i}, \ta_{\q,j}] = i \delta_{\p,-\q} (K(-\p)^{-1})_{ij} = -\frac{2\pi i}{k} \epsilon_{ij}\,\delta_{\p,-\q}\, \frac{e^{ip_i} + e^{-ip_j}}{1+\cos(p_1+p_2)} \,.
\end{equation}
In the quantum theory, the physical Hilbert space is defined via the remaining constraints: the Gauss law \eqref{eq:Gauss} and the framing constraint \eqref{eq:staggered_constraint}. The Gauss law plays the familiar role as the generator of gauge transformations, which act trivially on the physical Hilbert space. As a result, gauge non-invariant operators are projected out from the theory. In momentum space, the Gauss law reads 
\begin{equation}
\tilde{\mathcal G}_{\p} = (1+e^{i(p_1+p_2)})\epsilon_{ij} (1-e^{ip_j})\ta_{\p,i} = 0\,, 
\end{equation}
which clearly commutes with the constraints $\Pi_{\p,i} = 0$ for $\p \in \P$ because precisely these modes are not constrained by the Gauss law. Hence these constraints are first-class and we can consistently set the canonical momenta for the $\P$ modes to zero.

When the canonical momentum $\Pi$ of some coordinate $X$ vanishes, only states which are translationally invariant in $X$ contribute. In other words, in such a situation the physical states can all be thought of as $\int dX\,\ket{X}$ where $\ket{X}$ are eigenstates of the operator $X$. This means that the expectation value of operators of the type $e^{i\alpha X}$ with $\alpha\in \mathbb R$ will vanish. 

A generic local operator $O_{\x}$ will contain the peculiar modes with $\p \in \P$. However it is easy to project out these modes by instead considering the sum $O_{\x}+O_{\x+\s}$. The Fourier modes of this operator are
\begin{equation}
\tilde O_{\p} (1+e^{ip_1+ip_2}) e^{i \p\cdot \x}
\end{equation}
which identically vanishes whenever $\p \in \P$. Moreover, the expectation value of any operator $e^{i O_{\x}}$ will generically vanish because of the peculiar modes, but $e^{iO_{\x}+iO_{\x+\s}}$ will not. 

\begin{figure}[h] 
   \centering 
\includegraphics[width=.7\textwidth]{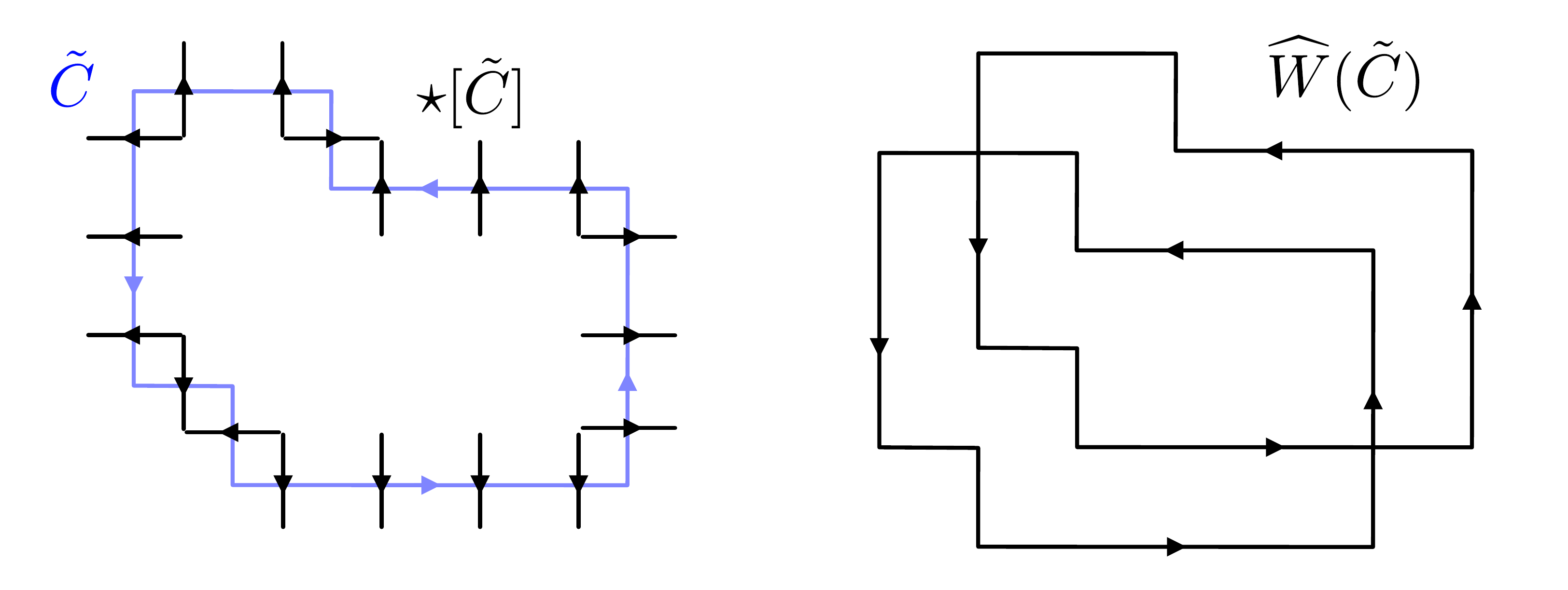} 
\caption{A framed Wilson line in the non-compact $\RR$ Chern-Simons theory. Here we show the curve $\tilde C$ on the dual lattice and its Poincar\'e dual $\star[\tilde C]$. }
\label{fig:framed}
\end{figure}

To see the connection to framing, let us apply this logic to Wilson loops. A naive Wilson loop $W(C)$ on a generic contour $C$ contains the $\P$ modes and is a vanishing operator on the Hilbert space! However, the \emph{framed} Wilson loop 
\begin{equation}
e^{i \frac{q}{2} \sum_{(\x,i) \in C} (a_{\x,i} + a_{\x+\s,i})}\,,
\end{equation}
does not contain the $\P$ modes and is a genuine operator in the theory. The framing constraint is so named because it projects out all unframed Wilson loops, leaving framed loops as the only physical operators (which, as we will see soon, are also topological).  We can write a general framed Wilson loop in a simple way using cup products. Let $\tilde C$ be a contour on the dual lattice, with $(\star[\tilde C])_\ell$ its Poincar\'e dual 1-cochain which is equal to the number of oriented times $\tilde C$ crosses the link $\ell$. The framed Wilson loop is simply
\begin{equation} \label{eq:R_wilson}
\W_q(\tilde C) = e^{i \frac{q}{2} \sum \star[\tilde C] \cup a - a \cup \star[\tilde C]}\,,
\end{equation}
where the sum is over all plaquettes of the lattice. Since we are in the non-compact $\RR$ gauge theory, $q$ can be an arbitrary real charge. See Fig.~\ref{fig:framed} for an example of a framed Wilson line. 

\begin{figure}[h] 
   \centering 
\includegraphics[width=.8\textwidth]{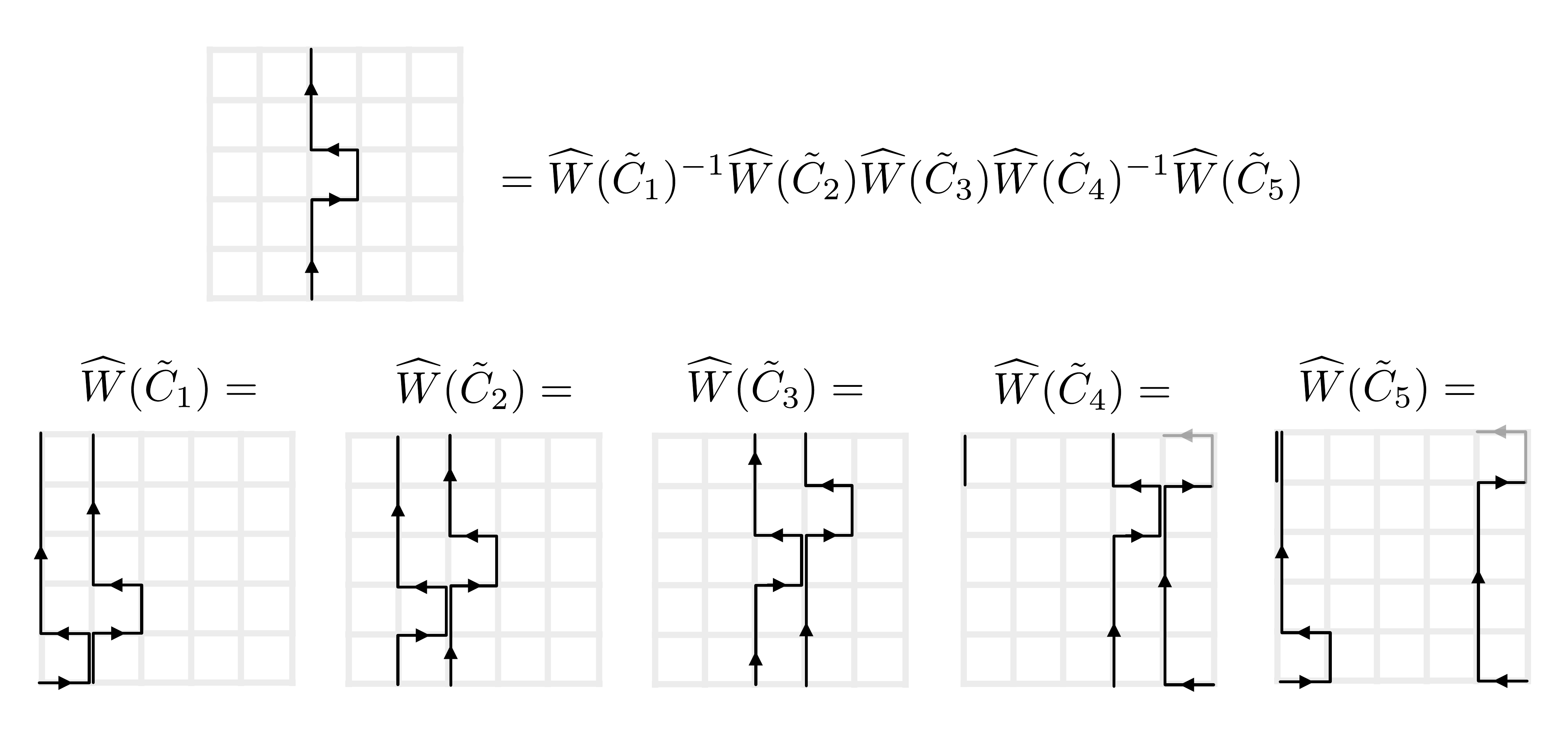} 
\caption{On a torus with odd side lengths $L_1, L_2$, any `unframed' Wilson line can be written as a product of $\lcm(L_1,L_2)$ framed Wilson lines displaced by $\s = \hat 1 + \hat 2$. In the above example, $L_1 = L_2 = 5$. }
\label{fig:unframed}
\end{figure}

The number of vanishing canonical momenta depends sensitively on the lengths of the torus $L_1, L_2$, and one may wonder whether the requirement that Wilson loops be framed also depends on the torus geometry. Such dependence would be rather unphysical. The more precise statement is that the framing constraint sets to zero any unframed Wilson loop \emph{which cannot be written in terms of framed Wilson loops}. For instance, if $L_1$ and $L_2$ are odd, $\P$ is empty, and there are no vanishing canonical momenta --- correspondingly, there are ordinary Wilson loops in this theory. The crucial point is that in this case, each ordinary unframed Wilson line can be written as a product of framed Wilson lines. We show an example in Fig.~\ref{fig:unframed}. The upshot is that regardless of the geometry of the torus, the physical operators in the theory are framed Wilson loops, with the framing dictated by our convention for the cup products in Eq.~\eqref{eq:action}. 

Let us now return to position space and write down the commutation relations between link fields. The commutator between two links is in fact non-local. But the commutator of a single link with a \emph{pair} of links shifted by $\s$ is ultra-local and given by
\begin{equation}\label{eq:basic_commutator} 
\begin{split}
[a_{\x,i}, a_{\y,j}+a_{\y+\s,j} ] &= \frac{1}{V} \sum_{\p,\q} [\ta_{\p,i},\ta_{\q,j}](1+e^{i(q_1+q_2)})e^{i\p\cdot \x + i \q\cdot \y} \\
&= -\frac{2\pi i}{k}\frac{1}{V}\sum_{\p} \epsilon_{ij} \frac{e^{ip_i} + e^{-ip_j}}{1+\cos(p_1+p_2)}(1+e^{-i(p_1+p_2)})e^{i\p\cdot(\x-\y)} \\
&= -\frac{4\pi i}{k}\frac{1}{V}\sum_{\p} \epsilon_{ij} \, e^{-ip_j} e^{i\p\cdot(\x-\y)} = -\frac{4\pi i}{k}\epsilon_{ij}\,\delta_{\x,\y+\hat j}\, .
\end{split}
\end{equation}
As a result, the commutator between two `point-split' pairs of links is given by 
\begin{equation}
[\frac{a_{\x,i}+ a_{\x+\s,i}}{2}, \frac{a_{\y,j}+ a_{\y+\s,j}}{2}] = -\frac{2\pi i}{k} \epsilon_{ij} \frac{\delta_{\x+\hat i, \y}+ \delta_{\x,\y+\hat j}}{2}\,,
\end{equation}
which is the lattice analogue of the continuum commutation relation
\begin{equation}
[a_i(\x),a_j(\y)] = -\frac{2\pi i}{k}\epsilon_{ij}\,\delta(\x-\y)\, .
\end{equation}
This leads to the following crucial formula involving an arbitrary pair of 1-cochains $X$ and $Y$:
\begin{equation} \label{eq:useful_equation} 
\left[\frac{i}{2}\sum_\Lambda ( X\cup a - a \cup X) \,, \; \frac{i}{2}\sum_\Lambda (Y\cup a - a \cup Y) \right] = \frac{2\pi i}{2k} \sum_\Lambda ( X \cup Y - Y \cup X)\, ,
\end{equation}
where the sum is over all plaquettes on the lattice. If we let $X = \star[\tilde C_1]$ and $Y = \star[\tilde C_2]$, the above commutation relation implies
\begin{multline}
\left[\frac{i}{2}\sum_\Lambda (\star[\tilde C_1] \cup a - a \cup \star[\tilde C_1])\,,\; \frac{i}{2}\sum_\Lambda (\star[\tilde C_2] \cup a - a \cup \star[\tilde C_2]) \right] \\
= \frac{2\pi i}{2k}\sum_\Lambda (\star[\tilde C_1] \cup \star [\tilde C_2] - \star[\tilde C_2] \cup \star [\tilde C_1]) = \frac{2\pi i}{k} \text{Int}(\tilde C_1,\tilde C_2). 
\end{multline}
As a result, framed Wilson loops obey the expected algebra
\begin{equation}
\W_{q_1}(\tilde C_1) \W_{q_2}(\tilde C_2) =  \W_{q_2}(\tilde C_2)\W_{q_1}(\tilde C_1) = e^{-\frac{2\pi i}{k} q_1q_2\text{Int}(\tilde C_1,\tilde C_2)}\,. 
\end{equation}

\begin{figure}[h!] 
   \centering
   \includegraphics[width=.7\textwidth]{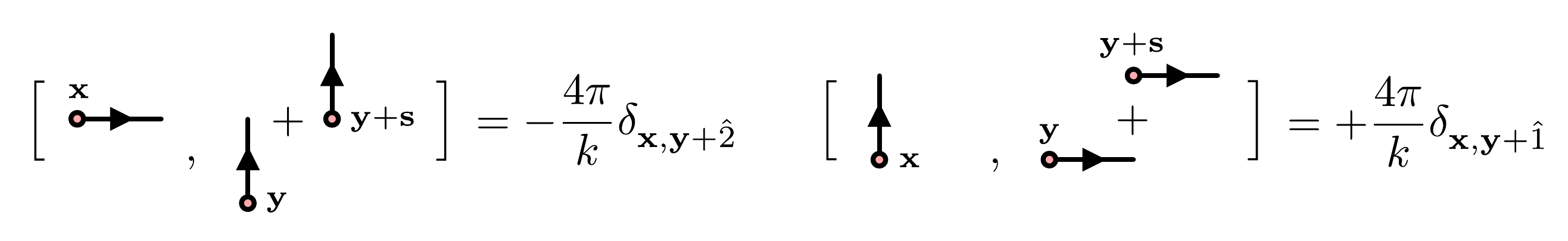} 
   \caption{ Commutation relations \eqref{eq:basic_commutator} between a single link field and a pair of links displaced by $\s = \hat 1 + \hat 2$. }
   \label{fig:commutation_1}
\end{figure}

\subsection{Symmetries and gauge redundancies}
\label{sec:R_redundancies} 

The action \eqref{eq:action} is invariant under shifts of the gauge field of the form $a\rightarrow a+\omega$ and $a_0\rightarrow a_0+\omega_0$ with the 1-cochain $\omega \in C^{1}(\mathbb R)$ and the 0-cochain $\omega_0\in C^{0}(\mathbb R)$ satisfying
\begin{equation}
d\omega=0\,, \ d\omega_0-\dot \omega=0\,.
\end{equation}
Now let us focus for the moment on the global transformation for which $\omega_0=0$ and $\omega$ is time-independent. Then we can associate $\omega \in Z^{1}(\mathbb R)$ and hence we can think of the Poincar\'e dual of $\omega$ as a (or some collection of) closed contour(s) $\tilde C$ on the dual lattice. Further we will assume that $\omega$ does not have Fourier modes in $\P$, because shifting these modes is automatically a symmetry since all the states must have a vanishing canonical momentum of these modes. 

The generator of this transformation is simply
\begin{equation}\label{eq:generator}
U[\omega]=e^{\frac{ik}{4\pi}\sum_{\x} \omega_{\x,i}(a_{\x+\hat i,j}+a_{\x-\hat j,j})\epsilon^{ij}}=e^{\frac{ik}{4\pi}\sum (\omega\cup a - a\cup \omega)}\,,
\end{equation}
which, since $d\omega = 0$, has the form of a framed Wilson loop. Specifically, comparing to Eq.~\eqref{eq:R_wilson}, $\W_q(\tilde C) = U\left[\frac{2\pi q}{k}\star[\tilde C]\right]$. Indeed this operator implements the shift of the gauge field 
\begin{equation}
U[\omega]^\dagger \, a_{\x,i}\, U[\omega] = a_{x,i} - \frac{i k}{4\pi}\sum_{\y,j,r}\omega_{\y,j}\, \epsilon^{jr}\,[a_{\y+\hat j,r}+a_{\y-\hat r,r},a_{\x,i}]=a_{\x,i}+\omega_{\x,i}\;,
\end{equation}
as is easily checked using Eq.~\eqref{eq:basic_commutator}. When $\omega = d\lambda$ is exact, $U[d\lambda]$ implements a gauge transformation and must be equivalent to the identity operator on the physical Hilbert space. Indeed, in this case 
\begin{equation} \label{eq:gauge_operator1} 
U[d\lambda] = e^{\frac{ik}{4\pi}\sum (d\lambda\cup a - a \cup d\lambda)} \equiv G[\lambda] =  e^{-\frac{ik}{4\pi}\sum (\lambda\cup da + da \cup \lambda)} = 1
\end{equation} 
thanks to the Gauss law \eqref{eq:Gauss}. A direct consequence is that a closed, contractible framed Wilson loop is equivalent to a generator of a gauge transformation, and hence is trivial. To see this, let $\lambda = \frac{2\pi q}{k}\star[\tilde D]$, so that $d\lambda = \frac{2\pi q}{k}\star[\partial\tilde D]$ where $\tilde D$ is a disk, and write
\begin{equation}
G\left[\frac{2\pi q}{k}\star[\tilde D] \right]= e^{-\frac{iq}{2}\sum (\star[\tilde D] \cup da - da \cup \star[\tilde D]) } = e^{\frac{iq}{2}\sum (\star[\partial\tilde D] \cup a - a \cup \star[\partial\tilde D]) } = \W_q(\partial\tilde D)\,,
\end{equation} 
where we used summation by parts. Therefore, contractible framed Wilson loops are trivial in the Hilbert space where the Gauss law is satisfied. 

The generators of the 1-form symmetry (which again are just framed Wilson loops) are indeed topological, since 
\begin{equation}
U[\omega] = G[\lambda]U[\omega] = e^{-\frac{2\pi i}{2k}\frac{1}{2}\sum d\lambda \cup \omega - \omega \cup d\lambda}\, U[\omega+d\lambda]  = U[\omega+d\lambda]\,,
\end{equation}
where in the first equality we used the Gauss law \eqref{eq:gauge_operator1}, the second equality follows from Eq.~\eqref{eq:useful_equation}, and in the last equality we integrated by parts and used $d\omega = 0$. In terms of the Wilson loop, this implies
\begin{equation}
\W_q(\tilde C) = G\left[\frac{2\pi q}{k}\star[\tilde D] \right] \W_q(\tilde C) = \W_q(\tilde C + \partial\tilde D).
\end{equation}
Intuitively, we can deform a framed Wilson loop by `tacking on' an arbitrary framed contractible loop, which is trivial thanks to the Gauss law. So Wilson loops only depend on their homology class. 

%

Now we consider another class of transformations which leaves the action invariant and is associated with the staggered modes $\P$. For this purpose, we take at least one of $L_1, L_2$ to be even. Consider a generic shift $a\rightarrow a+\omega$ with $\omega$ time-independent but not necessarily closed. The action changes by
\begin{equation}
\Delta S =-\frac{k}{4\pi} \int dt \sum_{\Lambda} \left(da_0\cup \omega-\omega\cup da_0\right)\;.
\end{equation}
We can choose $\omega$ to be a staggered in the following sense. Consider a fixed lattice site $x_0$ and take $\omega_{\x_0+n \s,i}=(-1)^n \, \varepsilon_i$ and all other $\omega_{\x,i} = 0$, where $n\in \mathbb Z$ and $\varepsilon_1,\varepsilon_2 \in \RR$. This shift is consistent with periodic boundary conditions as long as at least one of $L_1,L_2$ is even. It is easy to see that $X\cup \omega-\omega\cup X=0$ for \emph{any} 1-cochain $X$. Explicitly, the transformation is given by
\begin{equation}\label{eq:staggered_shift}
a_{\x,i}\rightarrow a_{\x,i}+\varepsilon_i \, \sum_{n=0}^{\lcm(L_1,L_2)-1} (-1)^n\, \delta_{\x,\x_0+n \s}\;,
\end{equation}
so that in Fourier components we have
\begin{equation}  
\begin{split}
\ta_{\p,i} &\rightarrow \ta_{\p,i}+\varepsilon_i\, \frac{1}{\sqrt{V}}\sum_{n=0}^{\lcm(L_1,L_2)-1} e^{i n(p_1+p_2-\pi)}e^{i  \p\cdot \x_0} \\
&= \ta_{\p,i}+\varepsilon_i\,e^{i  \p\cdot \x_0} \frac{\lcm(L_1,L_2)}{\sqrt{V}} \sum_{w\in\ZZ}\delta_{p_1+p_2,\pi + 2\pi w}
\end{split}
\end{equation}
The sum over $n$ forces the transformation to vanish unless $\p\in\P$. In other words, this symmetry only affects the zero modes discussed before. This transformation is implemented by the exponentiated canonical momentum of $\ta_{\p,i}$, with $\p \in \P$. But recall that any such canonical momentum vanishes \eqref{eq:staggered_constraint}, and hence all states in the Hilbert space are automatically invariant under the above transformation. In other words, the above transformation is the trivial identity operator on the Hilbert space. So the unusual staggered symmetry of the action is actually a gauge redundancy, not a global symmetry. In particular any operator which is not invariant under this symmetry (i.e. a generic unframed Wilson loop) would take a state out of the Hilbert space and is therefore forbidden. The canonical formulation makes it clear that symmetry of the lattice action that leads to the projecting out of unframed Wilson lines is in fact a gauge redundancy, as suggested in~\cite{Jacobson:2023cmr}.

\section{Compact Chern-Simons theory }
\label{sec:compact} 

We now want to construct the compact Chern-Simons theory with gauge group $U(1)$. This is achieved by gauging the $2\pi \ZZ$ subgroup of the $\RR$ 1-form symmetry of the non-compact theory. Intuitively, this turns the spatial gauge field into an angle-valued variable. For this purpose we introduce integer plaquette field $n_p$, and write a continuous-time action
\begin{equation}\label{eq:action1}
S=\frac{k}{4\pi} \int dt \sum_{\Lambda}\left[ a\cup \dot a -a_0\cup (da - 2\pi n)- (da-2\pi n)\cup a_0 \right]\,.
\end{equation}
Physically, $n_p$ can be thought of as the local magnetic flux, which can change by an exact integer under a large gauge transformation --- the sum of $n_p$ over all of space is gauge-invariant. We demand that $a\rightarrow a+2\pi m, n\rightarrow n+dm, m\in C^1(\mathbb Z)$ is a gauge symmetry (with $m$ time independent), as we will check in a moment. But first we check the ordinary (small) gauge transformation $a\rightarrow a+d\lambda$ and $a_0\rightarrow a_0+\dot \lambda$ are respected. Compared to the non-compact case \eqref{eq:small_gauge_variation}, we encounter additional terms  
\begin{equation}
\Delta S =\frac{k}{2}\int dt\sum_\Lambda  n\cup\dot\lambda + \dot\lambda \cup n\,.
\end{equation}
Since we take time to be non-compact we can integrate by parts in time and in order for the action to be invariant we must set $\dot n=0$. This is quite natural for an integer-valued field when time is continuous. Physically, this means that there are no monopole events which would change the quantized magnetic flux through space. To implement this constraint we follow the modified Villain procedure and introduce a Lagrange multiplier field $\varphi$ on the sites of the lattice, and add the term  
\begin{equation}
\int dt \sum_\Lambda \,  \dot\varphi \cup n
\end{equation}
to the action. When we quantize the theory, we impose that $\varphi$ is a periodic scalar $\varphi \sim \varphi + 2\pi$ so that its conjugate momentum, $n$, becomes an integer-spectrum operator. The operator $e^{i\varphi}$ is nothing but the monopole operator in our 2+1d $U(1)$ gauge theory. 

Now it is also possible to cancel the above gauge variation using a shift of $\varphi$. But this shift would involve a nonlocal shift of $\varphi$ by $\lambda$ at nearby sites. Instead we proceed as in Ref.\cite{Jacobson:2023cmr} and assign $\varphi$ charge $k$ under the gauge symmetry $\varphi \to \varphi - k\lambda$, while introducing an additional term
\begin{equation}
\frac{k}{2}\int dt \sum_\Lambda \, \dot a \cup_1 n\,.
\end{equation}
With these additional terms the variation of the action under small gauge transformations is
\begin{equation}
\Delta S = \frac{k}{2}\int dt\sum_\Lambda  n\cup\dot\lambda - \dot\lambda \cup n + d\dot\lambda \cup_1 n  = 0 
\end{equation}
using Eq.~\eqref{eq:cup_commutativity}. Now let us verify that $a \rightarrow a+2\pi m$ and $n\rightarrow n+dm$ (with $m$ time independent) is a symmetry:
\begin{equation}
\Delta S= \frac{k}{4\pi}\int dt\sum_\Lambda 2\pi m \cup \dot a + \int dt \sum_\Lambda \dot\varphi \cup dm \,.
\end{equation}
Both terms are total time derivatives, so the action is invariant. To summarize, the action
\begin{equation}\label{eq:cont_compact} 
S =  \int dt \sum_{\Lambda}\frac{k}{4\pi}\left[ a\cup \dot a -a_0\cup (da - 2\pi n)- (da-2\pi n)\cup a_0 + 2\pi \dot a \cup_1 n \right] + \dot\varphi\cup n 
\end{equation}
is invariant under the following transformations
\begin{equation}
\begin{split}
&a\rightarrow a+d\lambda+2\pi m\, \\
&n\rightarrow n+dm\,,\\
&\varphi\rightarrow \varphi-k\lambda\,,
\end{split}
\end{equation}
where $\lambda \in C^0(\RR)$ and $m \in C^1(\ZZ)$ and $\dot m = 0$. In fact, in the above discussion of the classical action we never used the fact that $m$ was integer valued, and correspondingly $k$ could still take any real value. This is due to the fact that we are taking time to be non-compact, so that gauge transformations cannot wind in time. As we will see below, even with time taken to be non-compact, level quantization and the discreteness of large gauge transformations emerge once we quantize the theory. 


\subsection{Quantization}

In conventional lattice notation, the full action in continuum time is given by
\begin{multline}
S=\int dt \sum_{\x}\Bigg[\frac{k}{4\pi}\left(a_{\x,i}\dot a_{\x+\hat i,j}\epsilon^{ij}-\left[(a_0)_{\x}+(a_0)_{\x+\s}\right]\left[ (da)_{\x,12}-2\pi n_{\x,12}\right]\right)\\+\left(\dot\varphi_{\x} -\frac{k}{2}(\dot a_{\x,2}+\dot a_{\x+\hat 2,1})\right)n_{\x,12}\Bigg]
\end{multline}
Again there is no dependence on  $\dot a_0$, so the canonical momentum of $a_0$ vanishes, which in turn implies the Gauss law (the normalization is for later convenience)
\begin{equation} \label{eq:Gauss2} 
\mathcal G_{\x} = \frac{k}{4\pi}\left[(da-2\pi n)_{\x-\s,12} + (da-2\pi n)_{\x,12} \right]=0\;.
\end{equation}
Switching to momentum space, i.e. 
\begin{align}
a_{\x,i}=\frac{1}{\sqrt{V}}\sum_{\p} \ta_{\p,i}\, e^{i\p\cdot\x}\;,\ n_{\x,12}=\frac{1}{\sqrt{V}}\sum_{\p} \tn_{\p}\, e^{i\p\cdot\x}\;,\ \varphi_{\x}=\frac{1}{\sqrt{V}}\sum_{\p} \tvarphi_{\p}\, e^{i\p\cdot\x}\;,
\end{align}
we get that the rest of the action can be written as
\begin{equation}
S=\sum_{\p} \left( \frac{1}{2}K(\p)^{ij}\, \ta_{\p,i}\, \dot \ta_{-\p,j}+ \tn_{\p} \left(\dot \tvarphi_{-\p}-\frac{k}{2}\dot \ta_{-\p,2}-\frac{k}{2}\dot \ta_{-\p,1}\, e^{-ip_2}\right)\right)\;.
\end{equation}
Defining $\overline{\tvarphi}_{\p}=\tvarphi_{\p}- \frac{k}{2}\ta_{\p,2}- \frac{k}{2}\ta_{\p,1}\, e^{ip_2}$ we write the above as
\begin{equation}
S=\sum_{\p} \left( \frac{1}{2}K(\p)^{ij}\, \ta_{\p,i}\dot \ta_{-\p,j} + \tn_{\p} \dot {\overline{\tvarphi}}_{-\p}\right)\;.
\end{equation}
This form of the action makes it clear that when $\p \not\in\P$ the operators $\ta_{\p,i}$ obey the commutation relations \eqref{eq:bcommutator} as before, and the canonical momenta $\Pi_{p,i}$ of modes $\ta_{\p,i}$ with $\p\in\P$ all vanish. On the other hand we have that 
\begin{equation}
[\overline{\tvarphi}_{\p},\tn_{\q}]=i\delta_{\p,-\q}\,.
\end{equation}
Now we assume that $\tn_{\p}$ commutes with $\ta_{\p,i}$ to get
\begin{equation}
[\tvarphi_{\p},\tn_{\q}]=i\delta_{\p,-\q} \ \Longrightarrow \ [\varphi_{\x}, n_{\y,12}]=i\delta_{\x,\y}\,.
\end{equation} 
So we can view $n$ as the canonical momentum for $\varphi$. But now notice that since 
\begin{equation}
[\ta_{\p,i},\overline{\tvarphi}_{\q}]=0 = [\ta_{\p,i},\tvarphi_{\q}-\frac{k}{2}\ta_{\q,2}-\frac{k}{2}\ta_{\q,1}\, e^{iq_2}] \,,
\end{equation}
$\tvarphi$ will have a non-trivial commutation relation with the gauge field (for $\p\not\in\P$)
\begin{equation}
[\tvarphi_{\p},\ta_{\q,i}]= i \pi\, \delta_{\p,-\q}
\frac{1+e^{i(p_1+p_2)}}{1+\cos(p_1+p_2)}(\delta_{i1}\, e^{-ip_1} - \delta_{i2})\,.\end{equation}
For instance, in position space this implies that
\begin{equation}
[\varphi_{\x}, a_{\y,i} + a_{\y+\s,i} ] = 2\pi i\, (\delta_{i1} \delta_{\x,\y+\hat 1} - \delta_{i2}\delta_{\x,\y})\,,
\end{equation}
so that
\begin{equation}
[\varphi_{\x}, (da)_{\y,12} + (da)_{\y+\s,12}] = 2\pi i\, (\delta_{\x,\y} - \delta_{\x,\y+\s})\,. 
\end{equation}
This leads to $[\varphi_{\x}, \mathcal G_{\y}] = -i k\, \delta_{\x,\y}$, so indeed the monopole operator $\mathcal M_{\x} = e^{i\varphi_{\x}}$ has electric charge $-k$~\cite{PhysRevD.34.3851,AFFLECK1989575,Jacobson:2023cmr}. In the following, we will also make use of the fact that
\begin{equation}
\sum_\Lambda\left( \varphi \cup X + \frac{k}{2}a\cup_1 X \right)
\end{equation}
commutes with all $a$, where here $X$ is an arbitrary 2-cochain.


\subsection{Gauge redundancies}
\label{sec:gauge_redundancies} 

Conventional 0-form gauge transformations $a\rightarrow a+d\lambda$, $\varphi \to \varphi - k\lambda$ are implemented by the analogous operator to 
\eqref{eq:gauge_operator1}, namely 
\begin{equation}\label{eq:gauge_operator2} 
G[\lambda] = e^{-\frac{ik}{4\pi}\sum \lambda \cup (da-2\pi n) + (da - 2\pi n)\cup \lambda}\,,
\end{equation}
which is equal to the identity operator on Hilbert space thanks to the Gauss law \eqref{eq:Gauss2}. An important consequence of the Gauss law is that the total magnetic flux through the torus necessarily vanishes.

We also want to impose compactness of $\varphi$, in other words we want to gauge discrete shifts $\varphi\to \varphi + 2\pi u$ with $u$ an independent integer on each site. This transformation is achieved by 
\begin{equation}
C[u] = e^{2\pi i \sum u \cup n}\,, \quad u \in C^0(\ZZ)\,. 
\end{equation}
Setting this operator equal to the identity on the physical Hilbert space implies that $n$ is an operator with an integer spectrum. 

Finally, we have the discrete gauge symmetry corresponding to large gauge transformations, which take $a\rightarrow a+2\pi m, n\rightarrow n+dm$ with  $m\in C^1(\mathbb Z)$. This is implemented by the operator
\begin{equation} \label{eq:large_gauge} 
U[m]=e^{i\sum \left[\frac{k}{2} (m\cup a-a\cup m + a \cup_1 dm)+\varphi\cup dm \right]}\,, \quad m\in C^1(\mathbb Z) \,.
\end{equation}
Now we must impose the constraint that $U[m]$ equals the identity operator on Hilbert space. For this to be consistent we also have to demand that the $U[m]$ operators are invariant under ordinary gauge transformations, but also commute between themselves, i.e. are themselves invariant under large gauge transformations. This requirement is what leads to the quantization of the level $k$ It is easy to verify that $U[m]$ and $G[\lambda]$ commute. However, in general the generators of large gauge transformations do not commute among themselves, but instead have the following commutation relations:
\begin{equation} \label{eq:large_gauge_commutation} 
 U[ m] \, U[ m'] = U[ m'] \, U[ m]\, e^{i k \pi \sum m\cup m' - m'\cup m} \,.
\end{equation}
For generic values of $k$, this non-commutativity prevents us from imposing the constraint $U[m] = 1$ on the Hilbert space, and can be thought of as a 't Hooft anomaly for the $2\pi \ZZ$ subgroup of the 1-form symmetry in the non-compact $\RR_k$ Chern-Simons theory. Clearly if $k$ is an even integer, the anomaly trivializes, and imposing the 1-form gauge constraint defines the physical Hilbert space of the $U(1)_k$ theory. 

It is well-known that when $k$ is an odd integer the $U(1)_k$ CS theory is a spin-TQFT~\cite{cmp/1104180750,Belov:2005ze}. In the current context, if $k$ is an odd integer one can introduce fermionic degrees of freedom on the lattice to cancel the unwanted phase in Eq.~\eqref{eq:large_gauge_commutation} in a procedure known as fermion condensation~\cite{Gaiotto:2015zta,Wan:2016php,Kapustin:2017jrc,Aasen:2017ubm}. For the moment we restrict ourselves to $k \in 2\ZZ$, and discuss the fermionic CS theory in Sec.~\ref{sec:fermionic_CS} below. 

In fact, to consistently gauge $2\pi$ shifts $a \to a + 2\pi m$ it is necessary but not sufficient to ensure that the generators $U[m]$ commute. To see the remaining issue, note that the operators as defined in Eq.~\eqref{eq:large_gauge} satisfy
\begin{equation}
U[m]\, U[m'] = U[m+m'] \, e^{i\frac{k\pi}{2} \sum m \cup m' - m' \cup m } \,.
\end{equation}
If $dm  = dm' = 0$, this phase is indeed trivial. Otherwise this phase can be nontrivial if $k$ is not a multiple of 4, and hence inconsistent with the constraint $U[m] =1$. We can fix this problem by rephasing the operators by $U[m] \to e^{i \frac{k\pi}{2}\sum m\cup m}\, U[m]$
so that 
\begin{equation} \label{eq:U_concatenation} 
U[m]\, U[m'] = U[m+m'] \, e^{i k\pi \sum m \cup m' }\,,
\end{equation}
which is innocuous if $k\in 2\ZZ$. 

\subsection{$\ZZ_k$ 1-form symmetry and 't Hooft anomaly}
\label{eq:Zk_anomaly} 

Recall that in the non-compact CS theory, framed Wilson loops with charge $q$ consisted of two charge-$q/2$ Wilson loops displaced by $\s$. In the compact CS theory, when $q$ is odd the two Wilson loops have fractional charges and have to be connected by a $\ZZ_2$ surface built out of the Villain variable $n_p$,
\begin{equation} \label{eq:compact_wilson} 
\W_q(\tilde C) = e^{\frac{iq}{2} \sum \star[\tilde C] \cup a - a \cup \star [\tilde C] + 2\pi \star[\tilde C] \cup_1 n}\,.
\end{equation}
Throughout the following we denote the minimally-charged Wilson line (which is now a well-defined notion) by $\W(\tilde C) \equiv \W_1(\tilde C)$. We can repeat the analysis in Sec.~\ref{sec:R_redundancies} to show that these framed spatial Wilson loops only depend on the homology class of $\tilde C$. To see this, let us use the 2d cup product identity from Footnote~\ref{fn:cup_identity}
\begin{equation}
\lambda \cup n - n\cup \lambda = d\lambda \cup_1 n, 
\end{equation}
to rewrite the generator of gauge transformations \eqref{eq:gauge_operator2} as
\begin{equation}
G[\lambda] = e^{\frac{ik}{4\pi}\sum (d\lambda \cup a - a \cup d\lambda) + 4\pi n\cup\lambda + 2\pi \, d\lambda \cup_1 n }\,.
\end{equation}
Now we set $\lambda = \frac{2\pi q}{k}\star[\tilde D]$, so that $d\lambda = \frac{2\pi q}{k}\star[\partial\tilde D]$ where $\tilde D$ is a disk and $q\in\ZZ$. Then one finds
\begin{equation}
G\left[\frac{2\pi q}{k}\star[\tilde D] \right]= e^{\frac{iq}{2}\sum (\star[\partial\tilde D] \cup a - a \cup \star[\partial\tilde D])+ 4\pi  n\cup\star[\tilde D] + 2\pi \star[\partial\tilde D] \cup_1 n } = \W_q(\partial\tilde D)\,,
\end{equation} 
so that contractible framed Wilson loops are pure gauge transformations and hence trivial, provided the Gauss law is satisfied. To see very explicitly that framed Wilson loops are topological, we can start with an arbitrary loop $\tilde C$ and add a contractible loop $\partial\tilde D$ using a gauge transformation
\begin{equation} \label{eq:deformation} 
\W(\tilde C) = G\left[\frac{2\pi}{k}\star[\tilde D] \right]\, \W(\tilde C) = \W(\tilde C + \partial\tilde D)\,,
\end{equation}
just as in the non-compact case. 

While contractible Wilson loops can be trivialized using a gauge transformation, non-contractible loops are necessarily non-trivial as they obey the familiar commutation relation
\begin{equation}
\W(\tilde C_1) \W(\tilde C_2) =  \W(\tilde C_2)\W(\tilde C_1) = e^{-\frac{2\pi i}{k} \text{Int}(\tilde C_1,\tilde C_2)}\,. 
\end{equation}
Moreover, in contrast to the non-compact case, we have that
\begin{equation}
[\W(\tilde C)]^k = \W_k(\tilde C) = e^{\frac{ik}{2} \sum \star[\tilde C] \cup a - a \cup \star [\tilde C] + 2\pi \star[\tilde C] \cup_1 n} = U[\star[\tilde C]] = 1
\end{equation}
by the large-gauge constraint (we remind the reader that in the current discussion we are taking $k$ to be even). So while contractible Wilson loops are trivial, the eigenvalues of the minimal non-contractible Wilson loops are non-trivial $k$th roots of unity.   

In particular, just as in the continuum, these relations imply the $k$-fold degeneracy of the Hilbert space on the torus. Of course, this commutation relation can be viewed as a central extension of the $\ZZ_k \times \ZZ_k$ symmetry acting on the Wilson loops winding around each cycle of the torus --- the hallmark of the 't Hooft anomaly of the $\ZZ_k^{(1)}$ 1-form symmetry generated by the Wilson loops themselves.

We emphasize that although the ingredients of our lattice model are bosonic, each carrying an infinite dimensional Hilbert space, the actual Hilbert space of the theory is finite and $k$-dimensional on the torus. While the non-trivial algebra obeyed by Wilson loops implies that all states in the Hilbert space are \emph{at least} $k$-fold degenerate, the fact that the only non-trivial operators in the theory are the topological framed Wilson loops means that the Hilbert space is finite dimensional. Just as in the continuum, the finiteness of the Hilbert space is spoiled if we include a conventional lattice Maxwell term, which makes the Hamiltonian non-zero and reinstates the unframed Wilson lines as physical and non-topological operators.

\subsection{Defect Hilbert spaces and topological spin}
\label{sec:defects} 

Up until this point, we have always been working in the Hilbert space where the local Gauss law \eqref{eq:Gauss2} and the framing constraint are both satisfied. While this gives rise to a well-defined Hilbert space, we can also access other superselection sectors where the constraints are violated in a controlled way. Such modifications of the contraints give rise to `defect' Hilbert spaces. 

In the case of the Gauss law, the defect Hilbert space corresponds to inserting temporal Wilson lines which represent the static worldlines of electric probe particles. If at least one of $L_1,L_2$ is even, the Lagrangian in Eq.~\eqref{eq:cont_compact} is invariant under staggered shifts $(a_0)_{\x} \to (a_0)_{\x} + \omega \sum_{n=0}^{\lcm(L_1,L_2)-1} (-1)^n \delta_{\x,\x_0+n\s}$, suggesting that we should only consider \emph{framed} temporal Wilson lines. Indeed, suppose we insert an ordinary, unframed temporal Wilson line of charge $q$ at a site at $\y$. This modifies the Gauss law to 
\begin{equation}
(da -2\pi n)_{\x-\s,12} + (da-2\pi n)_{\x,12} \stackrel{?}{=} \frac{4\pi q}{k}\delta_{\x,\y}\,.
\end{equation}
This constraint would be inconsistent --- in momentum space, the left-hand-side does not involve the $\P$ modes, while the right-hand-side does. To get a sensible defect, in general we have to split a charge-$q$ temporal Wilson line into a staggered pair of charge-$q/2$ lines, which gives rise to the modified Gauss law  
\begin{equation} \label{eq:defect_gauss} 
(da -2\pi n)_{\x-\s,12} + (da-2\pi n)_{\x,12} = \frac{2\pi q}{k}(\delta_{\x,\tilde \x_0-\frac{\s}{2} } + \delta_{\x,\tilde \x_0 + \frac{\s}{2}})\,,
\end{equation}
where $\tilde \x_0$ is a point on the dual lattice corresponding to the midpoint of the two charge-$q/2$ probes. 

The most general defect Hilbert space, which we denote by $\mathcal H_{\{(q_i, \tilde \x_i)\}}$ is labelled by the charges $q$ and positions $\tilde \x$ of the static probe particles. The defects are topological, corresponding to the fact that the framed temporal Wilson lines are nothing but the defects associated to the $\ZZ_k$ 1-form symmetry. Explicitly, given a state $|\psi\rangle \in \mathcal H_{(q,\tilde \x)}$ we have the unitary transformation 
\begin{equation}
|\psi\rangle \to \W_q(\tilde C_{\tilde \x\tilde \y}) |\psi\rangle\,,
\end{equation}
which maps $\mathcal H_{(q,\tilde \x)} \to \mathcal H_{(q,\tilde \y)}$. Here $\tilde C_{\tilde \x, \tilde \y}$ is any curve on the dual lattice connecting $\tilde \x$ and $\tilde \y$. The dependence on this curve is also topological, up to phases which can appear if the curve crosses other defects --- this is related to the anomaly. In general, we can connect two defect Hilbert spaces $\mathcal H_{\{(q_i,\tilde \x_i)\}}$ and $\mathcal H_{\{(q_i', \tilde \x_i')\}}$ if and only if $(\sum q_i) = (\sum q_i') \text{ mod } k$, so that they have the same charge under the 1-form symmetry. Indeed, the defect Hilbert space with a charge-$k$ temporal Wilson line at $\tilde \x$ can be mapped to the ordinary Hilbert space by acting with the local monopole operator $e^{-i  \bar\varphi_{\tilde \x - \frac{\s}{2}}}$, reflecting the fact that the 1-form symmetry is $\ZZ_k$. 

\begin{figure}[h!] 
   \centering
   \includegraphics[width=.7\textwidth]{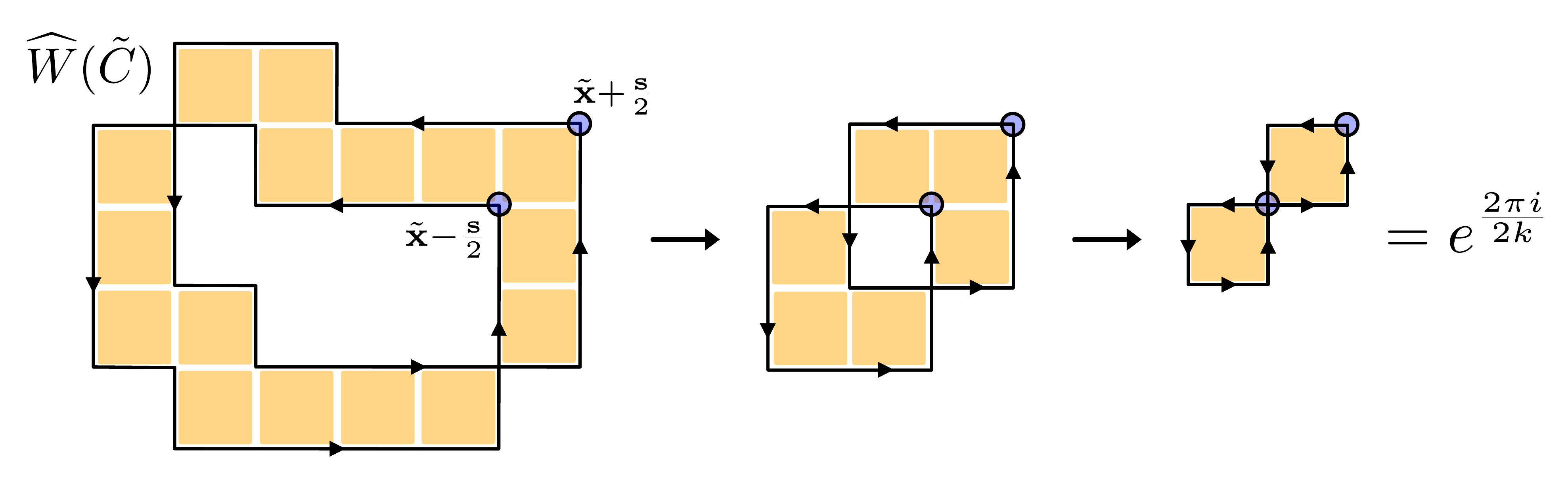} 
   \caption{ Shrinking a contractible Wilson loop in the presence of a temporal Wilson line yields a $\ZZ_{2k}$ phase which we interpret as the topological spin. The pair of circular blue dots denote the two charge-$1/2$ edges of the framed temporal Wilson line. }
   \label{fig:topological_spin}
\end{figure}

We can now compute a correlation function which we interpret as giving the topological spin. We consider the expectation value of a framed Wilson loop on a contractible curve $\tilde C$, but in the defect Hilbert space $\mathcal H_{(1,\tilde \x)}$ where $\tilde \x$ and $\tilde C$ are depicted in Fig.~\ref{fig:topological_spin}. We can use the argument near Eq.~\eqref{eq:deformation} above to deform the Wilson loop to a single (staggered) plaquette, 
\begin{equation}
\langle \W(\tilde C)\rangle_{(1,\tilde \x)} = \langle e^{\frac{i}{2}\left( (da-2\pi n)_{\tilde \x-\frac{3\s}{2}} + (da-2\pi n)_{\tilde \x - \frac{\s}{2}}\right)}\rangle_{(1,\tilde \x)} = e^{\frac{2\pi i}{2k} }
\end{equation} 
using the modified Gauss constraint \eqref{eq:defect_gauss}. Alternatively, we can  arrive at the same phase by applying an open spatial Wilson line to move the defect away from the loop $\tilde C$, shrinking the loop, and moving the defect back to its original location. 

We can also consider a defect Hilbert space corresponding to a fixed violation of the staggered constraint, for instance by setting $\Pi_{\p,i} = c_{\p,i}$ where $c_{\p,i}$ are some c-numbers and $\p \in\P$. Note that this does not define a \emph{local} defect Hilbert space, as the constraint is modified locally in momentum space. These defect Hilbert spaces correspond to superselection sectors with fixed eigenvalues under (linear combinations of) the transformations discussed near Eq.~\eqref{eq:staggered_shift}, and are mixed by acting with ordinary, unframed spatial Wilson loops. Breaking the staggered gauge redundancy explicitly by turning on a Maxwell term, or coupling to an electrically charged particle with a standard hopping term, will lead to dynamical mixing of these superselection sectors, and result in an infinite-dimensional Hilbert space, as discussed in Sec.~\ref{eq:Zk_anomaly}.

\section{Odd levels: the fermionic Chern-Simons theory}
\label{sec:fermionic_CS} 

In Sec.~\ref{sec:gauge_redundancies} we observed that when $k$ is odd the (naive) generators of large gauge transformations do not commute, but instead satisfy
\begin{equation} \label{eq:odd_k_anomaly} 
U[m] \, U[m'] = U[m+m'] \, e^{i k\pi \sum m\cup m'} = U[m']\, U[m] \, e^{i k\pi \sum m \cup m' + m' \cup m} \, .
\end{equation}
These non-trivial phases constitute an obstruction to setting $U[m]=1$ to define the physical gauge-invariant Hilbert space, i.e. a 't Hooft anomaly for the $2\pi \ZZ$ subgroup of the $\RR$ 1-form symmetry of the $\RR_k$ theory.\footnote{Alternatively, one can start with $\RR_k$ and gauge the non-anomalous $4\pi \ZZ$ subgroup of the 1-form symmetry to land on the $U(1)_{4k}$ theory. The $\ZZ_2$ subgroup of the $\ZZ_{4k}$ 1-form symmetry of $U(1)_{4k}$ has an 't Hooft anomaly, with the generators obeying \eqref{eq:odd_k_anomaly}.} 

An intuitive explanation for why there is an anomaly is as follows. Via Poincar\'e duality we can view the generators of large gauge transformations as operators supported on (not necessarily closed) lines. In fact the operator $U[m]$ is just a charge-$k$ framed Wilson loop, but \emph{without} the surface connecting the two edges of the ribbon (the last term in Eq.~\eqref{eq:compact_wilson}). When $k$ is odd, the Wilson lines at the edges of the ribbon have fractional charge and are not invariant under large gauge transformations, leading to the anomaly. One can attempt to fix this by including Villain fields in the interior of the ribbon, but the resulting operator acts on $\varphi$ and is therefore not the correct generator of large gauge transformations. 

We can cancel the anomaly by finding a suitable set of operators $S[m]$ which satisfy the same relations as Eq.~\eqref{eq:odd_k_anomaly}, so that we can define an improved generator of large gauge transformations $U[m]\, S[m]$ which can be consistently constrained to the identity on Hilbert space. This is only possible if we first enlarge our Hilbert space and the algebra of operators. There is not a unique choice for these additional degrees of freedom --- we will follow Ref.~\cite{Bhardwaj:2016clt} and build $S[m]$ out of a set of fermionic operators. The values of $S[m]$ on cohomology classes $m \in H^1(\ZZ)$ turn out to be related to the spin structure on the spatial torus. This is how the odd $k$ theory, which is a spin-TQFT, depends on spin structure. The same operators we discuss below were used in Refs.~\cite{Chen:2017fvr,Chen:2018nog,Chen:2019wlx} to define generalized Jordan-Wigner bosonization maps on the lattice.

To proceed, we introduce a pair of Majorana fermions $\gamma_p, \gamma_p'$ on plaquettes, with the usual algebra
\begin{equation}
\{\gamma_p, \gamma_{p'} \} = \{\gamma_p', \gamma_{p'}'\} = 2\delta_{p,p'}\,,\ \{\gamma_p, \gamma_{p'}'\} = 0\,. 
\end{equation}
The full Hilbert space is now enlarged to include the $2^N$-dimensional fermionic Hilbert space --- in a moment we will project down to a subspace which has the same dimension of the bosonic Hilbert space we started with. One can think of these Majoranas as the real and imaginary parts of the creation and annihilation operators 
\begin{equation}
c_p = \frac{1}{2}(\gamma_p + i \gamma_p'), \ c_p^\dagger = \frac{1}{2}(\gamma_p - i \gamma_p')\,.
\end{equation}
We define fermion parity on each plaquette to be $(-1)^{F_p} = i\gamma_p'\gamma_p$. 

The crucial building blocks for defining the fermionic CS theory are a set of link operators, schematically defined as $S_\ell = i \gamma_p \gamma_{p'}'$ where $p$ and $p'$ are the (positively oriented) plaquettes lying in the coboundary of $\ell$. More specifically, these operators are defined for positively-oriented links, 
\begin{equation}
S_{\x,1} = i \gamma_{\x,12}\, \gamma'_{\x-\hat 2,12}\,, \quad S_{\x,2} = i \gamma_{\x-\hat 1,12}\, \gamma'_{\x,12}\,,
\end{equation}
and extended to negatively oriented links via $S_{-\ell} = - S_{\ell}$, just as for a 1-cochain. These operators square to 1 and satisfy~\cite{Bhardwaj:2016clt,Chen:2017fvr,Chen:2019wlx}
\begin{equation}
S_\ell S_{\ell'} = S_{\ell'}S_\ell \, e^{i \pi \sum [\ell]\cup[\ell'] + [\ell'] \cup[\ell]}\,. 
\end{equation}
Now let $m \in C^1(\ZZ)$ and define~\cite{Bhardwaj:2016clt}
\begin{equation}
S[m] = e^{i \pi \sum_{\ell < \ell' \in \lambda} [\ell] \cup [\ell'] } \prod_{\ell\in\lambda}  S_\ell \,,
\end{equation}
where $\lambda$ is the ordered set of all links $\ell$ for which $m_\ell \not = 0$ mod $2$, and the product is taken from left to right in the order dictated by $\lambda$. The ordering-dependence of the two factors compensate so that the operator as a whole is independent of the choice of ordering. If $m= dq$ is exact, then 
\begin{equation}
S[dq] = \prod_p (i\gamma_p'\gamma_p)^{dq \cup_1 [p]} \equiv e^{i \pi \sum dq \cup_1 F}
\end{equation}
where $F_p$ is the fermion number (well-defined mod 2) of a given plaquette. If $dm =0$ but is not exact, then $S[m] = -  e^{i \pi \sum m \cup_1 F}$. Finally, if $dm \not = 0$ mod $2$, then 
\begin{equation}
S[m] = e^{i \pi \sum m \cup_1 F} \prod_p \gamma_p^{|(dm)_p|}
\end{equation}
with some fixed ordering of the product over $\gamma$'s. Note that these operators satisfy~\cite{Bhardwaj:2016clt}
\begin{equation} \label{eq:S_algebra} 
S[m]\, S[m'] = S[m+m']\, e^{i \pi \sum m \cup m'} = S[m']\, S[m]\, e^{i \pi \sum m\cup m' + m' \cup m} \,,
\end{equation}
reproducing the algebra Eq.~\eqref{eq:odd_k_anomaly}. 

To see how the spin structure comes into play, note that we are free to consider a family of operators 
\begin{equation}
S_\Gamma[m] = e^{i\pi \sum_\Gamma m} \, S[m]\,, 
\end{equation}
where the additional phase factor depends on a closed curve $\Gamma$ on the lattice, $\partial\Gamma = 0$. This phase factor preserves the algebra \eqref{eq:S_algebra}, and if $dm=0$, the operator above is just a sign times $e^{i\pi \sum m \cup_1 F}$. Let $m = \star[\tilde C]$ with $\partial\tilde C = 0$. Specifically, we find
\begin{equation} \label{eq:S_closed}
S_\Gamma[\star[\tilde C]] = e^{i\pi \sum \star[\tilde C] \cup_1 F} \begin{cases}
(+1)\, e^{i \pi \, \text{Int}(\Gamma, \, \tilde C)} = +1 & \text{ if $\tilde C$ is contractible,} \\
(-1)\, e^{i \pi \, \text{Int}(\Gamma, \, \tilde C)} & \text{ if $\tilde C$ is non-contractible.}
\end{cases}
\end{equation}
Now we would like to interpret these phases in terms of the spin structure on the torus. Following e.g. Ref.~\cite{10.1112/jlms/s2-22.2.365,Karch:2019lnn}, given a spin structure $\rho$ on the torus there is an associated quadratic form $q_\rho$ on $H_1(\ZZ_2)$ which is equal to 0 on cycles with anti-periodic boundary conditions and 1 on cycles with periodic boundary conditions. This form satisfies $q_\rho(C_1 + C_2) = q_\rho(C_1)+q_\rho(C_2) + \text{Int}(C_1,C_2)$. We can identify $e^{i\pi \sum \star[\tilde C] \cup_1 F}\, S_\Gamma[ \star[\tilde C]] = e^{i \pi q_\rho(\tilde C)}$, where the choice of $\Gamma$ is equivalent to a choice $\rho$ of spin structure. For this discussion only the $\ZZ_2$ homology of $\Gamma$ is important. Denoting the two cycles of the torus by $a,b$, we find  

\begin{equation}
\begin{tabular}{c|| c | c | c | c}
$\rho$ & AP/AP & AP/P & P/AP & P/P \\
\hline
$\Gamma$ & $a+b$ & $b$ & $a$ & $0$ 
\end{tabular}
\end{equation}
In short, the choice of $\Gamma$ is simply telling us which generating cycles of the torus have anti-periodic boundary conditions. Finally, if $dm \not=0$, the operator $S_\Gamma[m]$ is equal to $e^{i\pi \sum m \cup_1 F} \prod_p \gamma_p^{|(dm)_p|}$, up to a sign.

As described in the beginning of this section, the idea is to replace the constraint $U[m] = 1$, which is inconsistent when $k$ is odd, with $U[m]\, S_\Gamma[m] = 1$. When $dm=0$ mod $2$ this sets $U[m]$ equal to a sign times $(-1)^{\sum m \cup_1 F}$, but when $dm \not = 0$ mod $2$ this equates the open Wilson line (which acts on the bosonic Hilbert space) with a pair of Majorana operators at the endpoints (which act on the fermionic Hilbert space):
\begin{equation}
U[m] = e^{i \frac{k\pi}{2}\sum m\cup m} \, e^{i \sum  \frac{k}{2}(m\cup a - a \cup m+a\cup_1 dm) + \varphi \cup dm } = e^{i\pi \sum_\Gamma m} \, e^{i\pi \sum m \cup_1 F} \, \prod_p \gamma_p^{|(dm)_p|}\,.  
\end{equation}
The right hand side changes the fermion number by one unit on the plaquettes where $(dm)_p = 1$ mod $2$, while the left hand side shifts $n \to n + dm$. For this to be consistent, we must identify the fermion parity on a plaquette with the eigenvalue of the Villain field mod $2$: 
\begin{equation} \label{eq:spin_charge}
(-1)^{n_p} = (-1)^{F_p}\,.
\end{equation}
The physical Hilbert space is defined by this spin-charge relation \eqref{eq:spin_charge} together with the large-gauge constraint 
\begin{equation} \label{eq:large_gauge_constraint} 
U[m]\, S_\Gamma[m] = 1\,.
\end{equation}
We stress that the Hilbert space is defined with reference to some fixed $\Gamma$. Recall that in order to introduce the fermionic operators $\gamma, \gamma'$ to cancel the anomaly, we had to enlarge our Hilbert space by a factor of $2^N$. Physically, however, we should have a $k$-dimensional Hilbert space on the torus. The point is that the identification \eqref{eq:spin_charge} picks out just a single fermionic state for each bosonic state in the large Hilbert space unconstrained by \eqref{eq:large_gauge_constraint}, so that the physical Hilbert space has dimension $k$. 

Now consider the implications of the above discussion for a charge-$k$ framed Wilson loop. Using the large-gauge constraint, the spin-charge relation, and Eq.~\eqref{eq:S_closed}, we find
\begin{equation}
\begin{split}
\W_k(\tilde C) = e^{ik\pi \sum \star[\tilde C]\cup_1 n}\, U[\star[\tilde C]] &= e^{ik\pi \sum \star[\tilde C]\cup_1 F}\, S_\Gamma[\star[\tilde C]] \\
&= \begin{cases}
 +1 & \text{ if $\tilde C$ is contractible,} \\
(-1)\, e^{i \pi \, \text{Int}(\Gamma, \, \tilde C)} & \text{ if $\tilde C$ is non-contractible.}
\end{cases}
\end{split}
\end{equation}
For any choice of $\Gamma$, charge-$k$ contractible Wilson loops are always equal to 1 (in the absence of defects), while winding charge-$k$ Wilson loops are a non-trivial sign depending on the spin structure: 
\begin{equation}
\W_k(\tilde C) =  (-1)\, e^{i \pi \, \text{Int}(\Gamma, \, \tilde C)}\,.
\end{equation}
In particular, the value of a charge-$k$ framed Wilson loop wrapped on a non-trivial cycle changes by a sign when we change the spin-structure. This matches the continuum discussions in Refs.~\cite{Seiberg:2016rsg,Seiberg:2016gmd}. 
 
\begin{figure}[h!] 
   \centering
   \includegraphics[width=.8\textwidth]{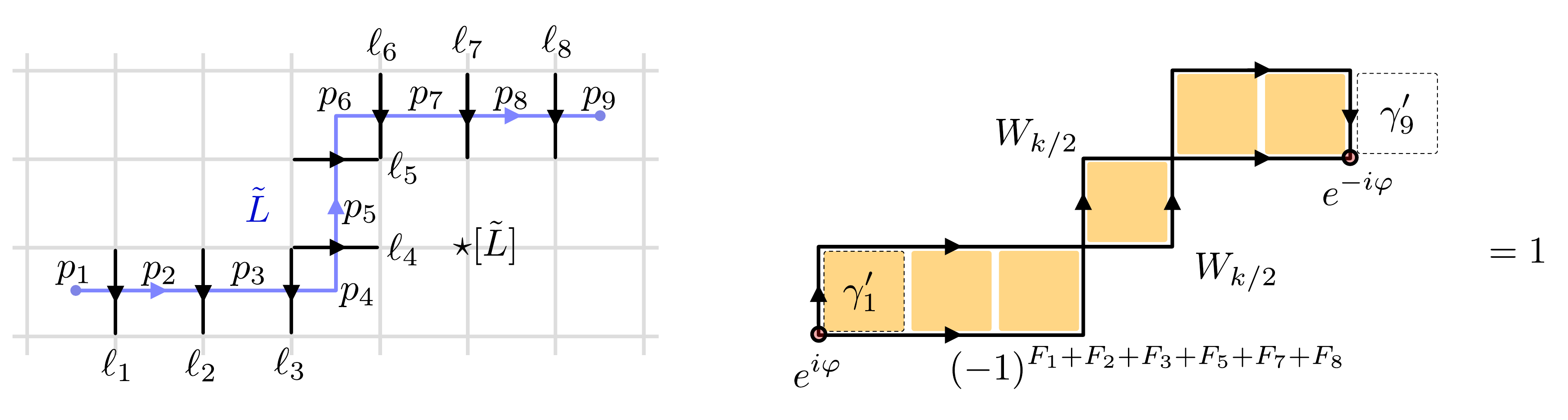} 
   \caption{ An open Wilson line equal to the identity operator by the large-gauge constraint.}
   \label{fig:open_line}
\end{figure}

Furthermore, the spin-charge relation tells us that the charge-$k$ temporal Wilson line represents the worldline of a fermion --- in the defect Hilbert space we can sum the modified Gauss law \eqref{eq:defect_gauss} over the entire lattice to find $\sum_{p} n_p = -1$, so that the net fermion parity is
\begin{equation}
(-1)^{\sum_p F_p} = -1\, .  
\end{equation}
The spin-charge relation also tells us that the monopole operator is a fermion --- $e^{i\varphi_{\x}}$ anti-commutes with $e^{i \pi n_{\x,12}}$, violating the spin-charge relation. This can be fixed by dressing the monopole with a fermion, e.g. $\gamma_{\x,12}'\, e^{i\varphi_{\x}}$. Of course, this operator is not gauge-invariant, and in the pure CS theory monopole operators have to be dressed by an open charge-$k$ Wilson line. In fact such an open Wilson line is nothing but the (trivial) operator $U[m] S_\Gamma[m]$. We show an example in Fig.~\ref{fig:open_line}. 

If we add electrically charged matter fields, we can define a genuine local monopole operator which is necessarily a fermion when $k$ is odd. In the continuum, such a fermionic monopole operator is nothing but the expected image of the Dirac fermion $\Psi$ in the basic bosonization map relating $\Psi$ to $U(1)_1$ coupled to a unit charge scalar~\cite{Wilczek:1981du,Polyakov:1988md,Shaji:1990is}.

\section{Conclusions}
\label{sec:conclusions}

In this work we have applied the Villain Hamiltonian approach to the canonical quantization of $U(1)_k$ Chern-Simons theory on a spatial lattice. Our discretization preserves defining properties of the continuum theory --- compactness of the gauge group, quantization of the level, the 1-form symmetry and its 't Hooft anomaly, the ground state degeneracy, and even its fermionic nature when the level is odd. More importantly, the lattice construction clarifies subtle aspects of the continuum theory. For instance, while in the continuum monopole operators are disorder operators, within the modified Villain approach on the lattice they have local expressions in terms of fields. The electric charges of these monopole operators arises naturally in our construction. Similarly, we identified how the framing of Wilson lines, which in the continuum is quite subtle, is incorporated on the lattice through constraints on the physical Hilbert space. As a result of this mechanism, the physical operators of the lattice theory are topological ribbons, whose algebra implies a $k$-fold degeneracy on the torus. 

The Hilbert space of our model is defined by a local gauge constraint (Gauss law), discrete gauge constraints imposing compactness of the $U(1)$ gauge group and of the (phase of the) monopole operator $\varphi$, and (depending on the details of the spatial torus) non-local constraints projecting out unframed Wilson loops. These constraints constitute a collection of `commuting projectors' which can in principle be used to define commuting projector Hamiltonians~\cite{Kitaev:1997wr,Levin:2004mi} where the various constraints are energetically imposed. In particular, when the number of sites in each direction is odd the non-local constraints disappear (since in that case ordinary Wilson lines are equivalent to products of framed ones), and the resulting commuting projector Hamiltonian will be local. The Hilbert space at each site would however be infinite-dimensional, and the resulting theory may end up being gapless --- see Refs.~\cite{Han:2021wsx,Han:2022cnr} for an in-depth study of similar issues in doubled CS theories. In addition, one may be able to use our construction to directly compute the topological entanglement entropy~\cite{Kitaev:2005dm,PhysRevLett.96.110405,Dong:2008ft} on the lattice, although the non-local constraints are likely to make this subtle.

One of the most interesting features of CS theory is its robust chiral edge modes~\cite{Elitzur:1989nr,Wen:1990se}. An obvious next step would be to analyze the edge modes of our lattice theory on spatial lattice with boundary, and verify that they are indeed chiral and define a sensible quantum field theory. We hope that this will give rise to new insights on the problem of discretizing genuinely chiral bosons and fermions on the lattice (see Refs.~\cite{Hellerman:2021fla,DeMarco:2023hoh,Berenstein:2023tru,Berkowitz:2023pnz,Berenstein:2023ric,Kaplan:2023pxd,Kaplan:2023pvd} for recent works on the subject). It is also interesting to go up in dimension, as we did in Ref.~\cite{Jacobson:2023cmr}, and study our CS theory as the boundary of a four-dimensional symmetry-protected topological (SPT) phase protected by the $\ZZ_k$ 1-form symmetry. Relatedly, leveraging the results in this paper one should be able to canonically quantize 4d $U(1)$ lattice gauge theory with a $\theta$ term to study the Witten effect, electric-magnetic duality, and more. 

Finally, one of our main motivations for studying the odd-level fermionic CS theories is to establish exact boson-fermion dualities on the lattice. While in this paper we focused on the pure CS theory, it is straightforward to introduce charged matter to obtain, for instance, $U(1)_1$ coupled to a scalar. Our analysis already correctly captures the fermionic statistics of the monopole in that theory (via the spin-charge relation \eqref{eq:spin_charge}), and constitutes a significant step towards proving a lattice-level duality with a Dirac fermion. We leave further exploration of this duality, and others, for future work.

\vspace{1cm}
\noindent\textbf{Acknowledgments} \\

\noindent We are grateful to P.S. Hsin for discussions and A. Cherman, T. Dumitrescu, Z. Komargodski, and S. Seifnashri for comments on the draft. 
T.J. is supported by a Julian Schwinger Fellowship from the Mani L. Bhaumik Institute for Theoretical Physics at UCLA. T.S. is supported by the Royal Society University Research Fellowship and in part by the STFC consolidated grant ST/T000708/1.

\begin{appendix}

\section{Quantizing a first-order Lagrangian}
\label{sec:first_order} 

Consider the Lagrangian
\begin{equation}
L= q_1\dot q_2\;.
\end{equation}
The canonical momenta are
\begin{equation}
\pi_1=0\;,\qquad \pi_2=q_1\;.
\end{equation}
The equations above are primary, second class constraints, which can be written in terms of the two functions
\begin{align}
f_1(\pi_1)=\pi_1\;,\\
f_2(\pi_2,q_1)=\pi_2-q_1\;.
\end{align}
The constrains are indeed second class as the Poisson bracket $[f_1,f_2]_{PB}=1$. Quantization then proceeds with defining the Dirac bracket as
\begin{equation}
[A,B]_{DB}=[A,B]_{PB}-[A,f_i]_{PB}(S^{-1})^{ij}[f_j,B]_{PB}\;,
\end{equation}
with $S_{ij}=[f_{i},f_{j}]=-(S^{-1})^{ij}=\begin{pmatrix}0&1\\-1 & 0\end{pmatrix}$ is an invertable matrix. Since we are imposing the constraint on $\pi_1$ and $\pi_2$, the only Dirac bracket we need is
\begin{equation}
[q_2,q_1]_{DB}=1\;.
\end{equation}
We quantize by promoting 
\begin{equation}
[q_2,q_1]_{DB}\rightarrow [q_2,q_1]=i
\end{equation}
where $[A,B]$ is the commutator of operators $A$ and $B$. The algebra then just simply becomes that $q_2$ is the conjugate momentum of $q_1$. 

However now notice that up to a total derivative, we can write the Lagrangian as
\begin{equation}
L= \frac{1}{2}(q_1\dot q_2-q_2\dot q_1)\;.
\end{equation}
The constraints are then given by
\begin{align}
f_1=\pi_1+\frac{1}{2}q_2\;,\\
f_2=\pi_2-\frac{1}{2}q_1\;.
\end{align}
The Poisson bracket between the two constraint again does not vanish
\begin{equation}
[f_1,f_2]_{PB}=-\frac{1}{2}[\pi_1,q_1]_{PB}+\frac{1}{2}[q_2,\pi_2]_{PB}=1\;.
\end{equation}
and so the constraint is second class.
The Dirac bracket is defined again in the same manner and gives the commutation relation
\begin{equation}
[q_2,q_1]=i
\end{equation}
i.e. the same as before. 

Let us explicitly compute the Dirac bracket in the first setup for two functions $A, B$ of $q_1,q_2,\pi_1,\pi_2$. Since 
\begin{equation}
[A,f_1]_{PB}= [A,\pi_1]_{PB}= \frac{\partial A}{\partial q_1}\;,\qquad [A,f_2]_{PB}=\frac{\partial A}{\partial q_2}+\frac{\partial A}{\partial \pi^1}\;,
\end{equation}
we easily get, by direct computation, that the Dirac bracket is given by
\begin{equation}
[A,B]_{DB}= \frac{\partial A}{\partial q_2}\left(\frac{\partial B}{\partial q_1}+\frac{\partial B}{\partial \pi_2}\right)-\frac{\partial B}{\partial q_2}\left(\frac{\partial A}{\partial q_1}+\frac{\partial A}{\partial \pi_2}\right)\;.
\end{equation}
Now, we view $A$ and $B$ as functions of $q_{1,2}$ and $\pi_{1,2}$. The derivative w.r.t. $\pi_1$ does not appear at all, so we can safely set $\pi_1=0$ in the dependence of $A,B$ before or after the Dirac bracket has been evaluated. However, we must be careful with setting $\pi_2=q_1$, and only impose this constraint after the Dirac bracket has been evaluated. In other words the Dirac bracket is given by
\begin{equation}
[A,B]_{DB}= \left[\frac{\partial A}{\partial q_2}\left(\frac{\partial B}{\partial q_1}+\frac{\partial B}{\partial \pi_2}\right)-\frac{\partial B}{\partial q_2}\left(\frac{\partial A}{\partial q_1}+\frac{\partial A}{\partial \pi_2}\right)\right]\Bigg|_{\pi_1=0,\pi_2=q_1}\;.
\end{equation}

Let $F(q_1,\pi_2)$ be a function of $q_1$ and $\pi_2$. The differential of $F$ is
\begin{equation}
dF=\frac{\partial F}{\partial q_1}dq_1+\frac{\partial F}{\partial \pi_2}d\pi_2\;.
\end{equation}
Now we want to set $q_1=\pi_2$ and look for a differential of $F(q_1,q_1)$. We find that it is given by
\begin{equation}
\frac{dF(q_1,q_1)}{dq_1}=\frac{\partial F}{\partial q_1}\Big|_{\pi_2=q_1}+\frac{\partial F}{\partial \pi_2}\Big|_{\pi_2=q_1}\;.
\end{equation}
Hence we get that setting the constraint $\pi_2=q_1$ before differentiation (which we denote by $A\big|$ and $B\big|$) renders the Dirac bracket between $A$ and $B$ to
\begin{equation}
[A,B]_{DB}= \frac{\partial A\big|}{\partial q_2}\frac{\partial B\big|}{\partial q_1}-\frac{\partial B\big|}{\partial q_2}\frac{\partial A\big|}{\partial q_1}\;.
\end{equation}
In other words, we can really just set $q_1$ to be the conjugate momentum of $q_2$.

This discussion can be directly applied to our lattice action \eqref{eq:action}. The momentum space action \eqref{eq:momentum_space_action} leads to the constraint on canonical momenta 
\begin{equation} \label{eq:constraint} 
\Phi_{\p,i} = \Pi_{\p,i} + \frac{1}{2}K(\p)_{ij}\, \ta_{-\p,j}\,.
\end{equation} 
The above constraint will have a non-vanishing Poisson bracket if $K(\p)_{ij} \not =0$, and constitutes a second-class constraint requiring the Dirac bracket. Proceeding with the modes for which $\p \not\in \P$. The Poisson bracket between the remaining constraints is
\begin{equation}
\begin{split}
F_{\p\p',ii'} \equiv [\Phi_{\p,i}, \Phi_{\p',i'}]_{\text{PB}} &=\frac{1}{2} K(\p)_{ij} \, [ \ta_{-\p,j}, \Pi_{\p',i'}]_{\text{PB}} +\frac{1}{2} K(\p')_{i'j'}\, [\Pi_{\p,i},\ta_{-\p',j'}]_{\text{PB}} \\
&= \frac{1}{2}\delta_{-\p,\p'}K(\p)_{ii'} - \frac{1}{2}\delta_{\p,-\p'}K(\p')_{i'i} = \delta_{\p,-\p'}K(\p)_{ii'}\,. 
\end{split}
\end{equation}
To consistently quantize the system we have to use Dirac brackets. The inverse of the matrix of constraints (continuing to assume $\p \not\in\P$) is
\begin{equation}
F^{-1}_{\p\p',ii'} = -\frac{1}{2}\left(\frac{4\pi}{k}\right)^2\frac{1}{1+\cos(p_1+p_2)}\delta_{\p,-\p'}K(-\p)_{ii'}\,,
\end{equation}
so the Dirac bracket between two momentum modes is 
\begin{equation}
\begin{split}
[\ta_{\p,i}, \ta_{\q,j}]_{\text{DB}} &= [\ta_{\p,i}, \ta_{\q,j}]_{\text{PB}} - [\ta_{\p,i}, \Phi_{\p',i'}]_{\text{PB}}\, F^{-1}_{\p'\q',i'j'}\, [\Phi_{\q',j'}, \ta_{\q,j}]_{\text{PB}} \\
&= 0 - (\delta_{\p,\p'}\delta_{ii'})\frac{1}{2}\left(\frac{4\pi}{k}\right)^2\frac{1}{1+\cos(p_1'+p_2')}\delta_{\p',-\q'}K(-\p')_{i'j'}\delta_{\q,\q'}\delta_{jj'} \\
&= -\frac{1}{2}\left(\frac{4\pi}{k}\right)^2\frac{1}{1+\cos(p_1+p_2)}\delta_{\p,-\q}K(-\p)_{ij} = \delta_{\p,-\q}(K(-\p)^{-1})_{ij}\,. 
\end{split}
\end{equation}
Upon quantization this reproduces Eq.~\eqref{eq:bcommutator}. In the above analysis we did not include the Gauss law constraint, which despite being second-class, with 
\begin{equation}
[\tilde{\mathcal G}_{\p}, \Phi_{\q,i}]_{\text{PB}} =(1+e^{i(p_1+p_2)})\delta_{\p,\q}\,\epsilon_{ij}(1-e^{ip_j})\,,
\end{equation}
only involves the coordinates and hence does not contribute to the Dirac bracket.

\end{appendix}

\newpage
\linespread{1}\selectfont
\bibliographystyle{utphys}
\bibliography{lattice_cs}

\end{document}